\documentclass[12pt]{article}
\usepackage{graphicx}
\begin{document}

\pagestyle{empty}

\noindent
{\bf Amplitude Variations in Pulsating Red Supergiants}

\bigskip

\noindent
{\bf John R. Percy and Viraja C. Khatu\\Department of Astronomy and Astrophysics\\University of Toronto\\Toronto ON\\Canada M5S 3H4}

\bigskip

{\bf Abstract}  We have used long-term AAVSO visual observations and Fourier 
and wavelet analysis to identify periods and to study long-term amplitude
variations in 44 red supergiants.  Of these, 12 stars had data which were
too sparse and/or had low-amplitude and/or were without conspicuous peaks in the Fourier
spectrum; 6 stars had only a long (2500-4000 days) period without significant
amplitude variation.  The other 26 stars had one or two periods, either
``short" (hundreds of days) or ``long" (thousands 
of days),
whose amplitudes
varied by up to a factor of 8, but more typically 2-4.  The median
timescale of the amplitude variation was 18 periods.  We interpret the
shorter periods as due to pulsation, and the longer periods as analogous
to the ``long secondary periods" found in pulsating red giants.

\bigskip

\noindent
{\bf 1. Introduction}

\smallskip

Red supergiant stars are extreme and complex objects.  Several dozen of
them have been intensively studied over the years, on account of their
brightness, complexity, and evolutionary importance.  Long-term spectroscopic 
and photometric
observations of individual objects such as Betelgeuse (Kervella {\it et al.} 2013) and
Antares (Pugh and Gray 2013a, 2013b) show complex variability on several
time scales, ranging from hundreds to thousands of days. 

Kiss {\it et al.} (2006) have used long-term visual observations from the
AAVSO International Database (AID) to study the variability of 46 pulsating
red supergiant stars.  They found periods in most of them, and two periods
in 18 stars.  The periods divide into two groups: periods of a few hundred
days which are probably due to pulsation, and periods of a few thousand
days which correspond to the ``long secondary periods" (LSPs) in pulsating red
giants; the nature of these LSPs is uncertain (Nicholls {\it et al.} 2009).  The shapes of
the peaks in the individual power spectra of these stars suggest that pulsation and
convection interact strongly.  The power spectra also exhibit strong
1/f noise, suggestive of irregular photometric variability caused by
large convection cells, analogous to the granulation seen
in the sun, but much larger.  Percy and Sato (2009) used self-correlation analysis to determine
or confirm LSPs in many of these same stars, using the same visual data.

Percy and Abachi (2013), hereinafter Paper 1, have recently used long-term visual
observations from the AID to study {\it amplitude variations} in
several samples of pulsating red {\it giants}: 29 single- and 30 double-mode SR
stars, 10 Mira stars, and the LSPs 
in 26 SR stars.  In each case, the amplitude varied
significantly, on a time scale (L) which was about 30-45 times the pulsation
period (P) or the LSP.  The fact that L/P is approximately constant for these
different groups of stars may be a clue to the origin of the amplitude variations.  It
suggests that either there is a causal relation between the two processes,
or they are both linked to some property of the stars, such as its radius.
In the present paper, we extend that study to 44 pulsating red
{\it supergiant} stars.  

\medskip

\noindent
{\bf 2. Data and Analysis}

\smallskip

We used visual observations, from the AAVSO International Database,
of the red supergiant variables listed in Table 1.  See "Notes on Individual
Stars" for remarks on some of these.  Our data extend a few years longer than that
of Kiss {\it et al.} (2006).

Paper 1 discussed some of the limitations of visual
data which must be kept in mind when analyzing the observations, and
interpreting the results.  The present study is even more challenging than
that of Paper 1, since the periods of red supergiants are
even longer than those of red giants, and the timescales of the amplitude
variations may be 20-40 times these periods.

The data, extending from JD(1) (as given in the table) to about JD 2456300 (except as noted in section 3.2), 
were analyzed using the
{\it VSTAR} package (Benn 2013; www.aavso.org/vstar-overview), especially the Fourier 
(DCDFT) analysis and wavelet (WWZ) analysis routines.  As noted by 
Kiss {\it et al.} (2006), the peaks in the power spectrum are complex,
rather than sharp, and it is not always possible to identify an exact period.

For the wavelet analysis, as in Paper 1, the default values were used for the decay time
c (0.001) and time division $\Delta$t (50 days).  The results are sensitive
to the former, but not to the latter.  Templeton, Mattei, and Willson (2005)
also used c = 0.001.  We used the DCDFT routine to inspect the period spectrum
of each star, but we were also guided by the results of Kiss {\it et al.} 
(2006) who determined periods for each star, and also listed other periods from
the literature.  For the WWZ analysis: around
each of the true periods, we generated the amplitude {\it versus} JD graph,
and determined the range in amplitude, and the number (N) of cycles of amplitude
increase and decrease.  As discussed in Paper 1, there was
often less than one cycle of amplitude variation, in which case our
estimate of N is a crude and possibly unreliable one.

\medskip

\noindent
{\bf 3. Results}

\smallskip

\noindent
3.1 Summary of Results

Interpreting the results is challenging 
because of the complexity of the stars, the low
amplitudes in many stars, and the long time scales involved.  Some stars
had several low-amplitude peaks in the DCDFT spectra which could not be shown to be significant,
unless there was also V data.  When the peaks were about a year,
and with amplitudes less than 0.1, they might be spurious results of the
Ceraski effect (see below).  The reality of very long periods was also uncertain if the
dataset was short and/or the star showed very long-term irregular variations.  In the
end, we have been conservative about which stars to include in Table 1, and
which to use to draw conclusions.

Table 1 lists the results for each star: the name, period P in days, initial JD,
amplitude range, number N, and length L in days of cycles of amplitude
variation, and the ratio L/P.  An asterisk in the table, or in the following
paragraphs indicates that there is a note about
the star in section 3.2.  Especially for the stars with N $\le$1, the amplitude
range is a lower limit.

The following stars were rejected as supergiants by Kiss {\it et al.} (2006):
UZ CMa, T Cet, IS Gem, Y Lyn, and are not listed in the table.

Almost all of the stars showed signals at a period of about one year, with an
amplitude of a few hundredths of a magnitude.  This was most likely due
to the Ceraski effect, a physiological effect of the visual observing
process.  Unless the signal was also present in V data (which would
not be subject to this effect), we assumed that the signal was
spurious.  It is possible, of course, that one or two of these stars
had a real low-amplitude period of about a year.

For the following stars: either the data were too sparse, and/or the
amplitude was too small, and/or the DCDFT spectrum showed no conspicuous
peaks: NO Aur*, CK Car*, IX Car*, W Cep, ST Cep*, AO Cru*, BI Cyg*,
WY Gem*, RV Hya*, XY Lyr*, 
AD Per*, PP Per*.

The following stars showed possible long (thousands of days) periods,
but without any significant variation in amplitude over the timespan of
the data: AZ Cyg* (3600 days), 
BU Gem (2500 days), RS Per* (4000 days), KK Per (3030 days),
PR Per* (3090 days), and FZ Per* (3440 days).

Figure 1 shows the amplitude variation for S Per.  Paper 1 shows the
amplitude variations for several pulsating red giants; these are similar to those
in the red supergiants.

\begin{figure}
\begin{center}
\includegraphics[height=8cm]{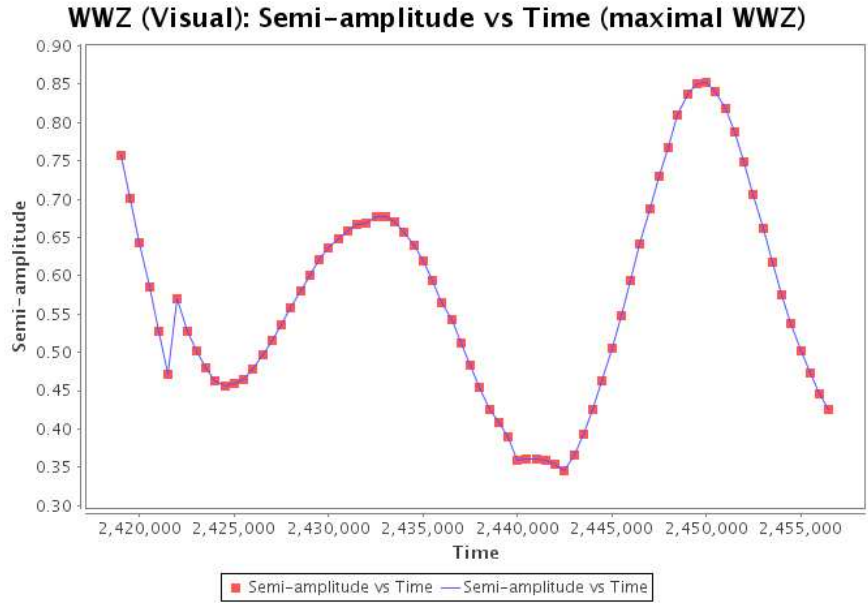}
\end{center}
\caption{The changing visual amplitude of the pulsating red supergiant S Per, with
a pulsation period of 813 days.  The amplitude varies between 0.35 and 0.85, and there
are approximately 2.25 cycles of increase and decrease.}
\end{figure}

\medskip

\noindent
3.2 Notes on Individual Stars

{\it SS And:}  In Table 1, we have given results using the literature
period of about 120 days, even though that period has an amplitude less than 0.03
in our data.  The highest peaks (1850$\pm$100 days) have amplitudes less than 0.05.

{\it NO Aur:} The star fades noticeably at the end of the (sparse) dataset.

{\it VY CMa:}  The long period is quite conspicuous in the light curve.
The DCDFT spectrum includes its aliases.

{\it RT Car:}  A period of about 400$\pm$ days is present in the V data, as
well as in the visual data.  The data are sparse after JD 2453000.

{\it BO Car:}  The 330-day period is sufficiently different from a year
that we have accepted it as probably real.

{\it CK Car:} There is a deep R CrB-like minimum at the start of the dataset,
and a possible shallower fading, halfway through the dataset.  This makes it
difficult to identify any underlying period.  There may or may not be a
period around 900-950 days.

{\it CL Car:}  There are comparable peaks at 500 and 1350 days which are
aliases of each other; the shorter period is the one which seems to be
present in the light curve, at least after JD 2452500.  The variability
before then is not well-defined.

{\it EV Car:}  The data are sparse.

{\it IX Car:}  There  is a gap in the data between JD 2441500-3500.  The light
curve shows a very slow rise and fall from beginning to end.  The DCDFT
spectrum shows several peaks between 2000-10000 days which may simply be a
mathematical way of representing the very slow variation.

{\it ST Cep:}  The light curve shows very slow variations; the DCDFT spectrum
shows several peaks between 2000 and 10000 days which may simply be a way of
mathematically representing the slow, probably-irregular variations.

{\it $\mu$ Cep:}  This is a well-observed (visually and photoelectrically)
star which is not excessively bright, and therefore much easier to observe
than $\alpha$ Ori or $\alpha$ Sco.

{\it AO Cru:} The 340-day period is close to a year, and of small amplitude;
it may well be spurious.

{\it AZ Cyg:}  The light curve shows slow, irregular variations; the DCDFT
spectrum shows several peaks between 2000 and 5500 days which may simply be
a way of mathematically representing the slow, probably-irregular variations.
There is no evidence for a significant period around 500 days.

{\it BI Cyg:}  There are many peaks in the range of hundreds and thousands
of days, but none stand out.

{\it TV Gem:} The 426 and 2550-day periods given by Kiss {\it et al.} (2006)
are aliases of each other.

{\it WY Gem:}  The data are dense, but there are no significant peaks with
amplitudes greater than about 0.05 mag.  This includes the 353-day period
reported by Kiss {\it et al.} (2006), which may be spurious.

{\it $\alpha$ Her:} The period of about 124 days, given by Kiss {\it et al.}
(2006) has an amplitude less than 0.03 in our data.  The highest peaks are
in the range 1300-1600 days, and the WWZ results suggest that the long period
is in the range 1500-1600 days.

{\it RV Hya:} The data are very sparse.

{\it W Ind:}  There is very little visual data after JD 2451500, but there is
extensive V data.  A period of 198 days is present in both datasets.

{\it XY Lyr:}  The literature period of 120-122 days has an amplitude less
than 0.03 mag in our data.  The highest peaks, 1850$\pm$100 days, have
amplitudes less than 0.05 mag.

{\it $\alpha$ Ori:}  This is the brightest and best-studied pulsating
red supergiant, though both visual and photoelectric observations are 
difficult because of the lack of convenient comparison stars.  A workshop
about this star has recently been held (Kervalla {\it et al.} 2013).
The period of 388 days (Kiss {\it et al.} 2006) is present in the V data,
but is not prominent in the visual data.

{\it S Per:}  The star fades by about a magnitude towards the end of the dataset.

{\it W Per:}  Our periods agree with the literature periods.  The 2875-day
period is clearly visible in the light curve.

{\it RS Per:}  There is a period of 224 days in the V data which does not
appear to be present in the visual data.

{\it XX Per:}  The literature periods are 4100 and 3150$\pm$1000 days; the
latter is not inconsistent with
our period of 2400 days.

{\it AD Per:} Peaks are at the long period of 3240 days, and at one year; the
latter is not present in the V data, and is probably spurious.

{\it BU Per:} There is a large gap in the middle of the dataset.  There are
peaks near a year, which are probably spurious.  In the table, we have given
results for the 381-day period (which may not be real).  There are peaks at 3700 and
5500 days (which are possibly aliases of each other); a time scale of about
5500 days appears to be present in the light curve.

{\it FZ Per:}  The V data do not support a period of about a year, which
is present in the visual data, and is almost certainly spurious.

{\it PP Per:} The only peak is at one year.

{\it PR Per:}  Peaks are at the long period of 3090 days, and at one year.
The amplitude of the long period increases only slightly.

{\it $\alpha$ Sco:}  Both visual and photoelectric observations of this bright
star are challenging, because of the lack of convenient comparison
stars.  The most recent long-term spectroscopic and photometric studies of
its variability are by Pugh and Gray (2013a, 2013b).  We find evidence, in the WWZ
analysis, for periods in the range of 1000-2000 days.

{\it W Tri:} The period of about 107 days, given by Kiss {\it et al.} (2006)
has an amplitude less than 0.03 in our data.  The highest peaks are at 595
and 765 days, both with amplitudes of about 0.07 mag.  The WWZ contours suggest 
that both of these periods are present, with variable amplitude, with N $\sim$ 1.0.

\medskip

\noindent
{\bf 4. Discussion}

\smallskip

All of the 26 stars in Table 1 show amplitude variations though, in some cases,
they are small -- a few hundredths of a magnitude.  They are, of course,
lower limits to what might be observed if the dataset was much longer.  The
timescales of the amplitude variation of the stars for which this timescale
is reasonably well-determined -- those with one or more cycles of increase
and decrease -- have a median value of 18 periods, with some uncertainty.  This value is similar for the
short periods and the long ones.
In the pulsating red giants (Paper 1), this ratio was 30-45, which is
significantly greater than for the supergiants.

Photometric observations of red supergiants in both the Milky Way and the
Magellanic Clouds suggest that the ``short" period is to be identified
with pulsation, and the ``long" period with the long secondary periods in
red giants.  Pugh and Gray (2013b) have identified an additional period
of 100 days in $\alpha$ Sco A (Antares) which is present in long-term spectroscopic observations, but
not in photometry.  They suggest that it is a convection-driven non-radial
p-mode.

Paper 1 raised the hypothesis that the amplitude variations might be
associated with the rotation of a star with large-scale convective regions.
Both simulations and observations show that red supergiants have such regions
on their surfaces (Chiavassa {\it et al.} 2010).
The approximate uniformity of the L/P values suggests that there is some link between
the period P and the process which causes the amplitude variations.
Stars
which are larger pulsate more slowly, and rotate more slowly, though the
exact relation between these two time scales depends on the distributation
of mass and of angular momentum in the star.
The rotation period of a star is 2$\pi$R(sin i)/(v sin i) where R is the
radius and (v sin i) is the measured projected equatorial rotation velocity.
Fadeyev (2012) gives the radii of SU Per and W Per as 780 and 620 solar
radii, respectively.  The rotation periods are therefore 
31000 (sin i)/(v sin i) and 39000 (sin i)/(v sin i) days, respectively.  The
measured values of (v sin i) are very uncertain; for Betelgeuse, the
value is probably a few km/s (Gray 2000, and private communication 2013).  
If this value also
applies to SU Per and W Per then, using an average value of sin i of 0.7,
the rotation periods are not inconsistent with the values of L, which are
21500 and 11900 days, respectively, with considerable uncertainty.

This paper also has an important education application: author VK is an
astronomical sciences major at the University of Toronto, and this is her
first formal research project -- a useful contribution to science, and an
interesting (we hope) example, for observers, of how their observations
contribute to both science and education.


\medskip

\noindent
{\bf 5. Conclusions}

\smallskip

We have identified one or two periods in 26 of 44 red supergiants, using
Fourier analysis of sustained, systematic visual observations.  All these
periods have amplitudes which vary by factors of up to 8 (but more typically
2-4) on time scales which are typically about 20 times the period.  The
``short" periods (hundreds of days) are presumably due to pulsation.  The
``long" periods are analogous to the long secondary periods in red giants;
their cause is not known for sure.  But there is observational and
theoretical evidence for large-scale convection in these stars, so a
combination of pulsation, convection, and rotation may combine to produce
the complex variations that are observed in these stars.

\medskip

\noindent
{\bf Acknowledgements}

\smallskip

We thank the hundreds of AAVSO observers who made the observations which were
used in this project, and we thank the AAVSO staff for processing and
archiving the measurements.  We also thank the team which developed the
{\it VSTAR} package, and made it user-friendly and publicly available.
This project
made use of the SIMBAD database, which is operated by CDS,
Strasbourg, France.

\medskip

\noindent
{\bf References}

\smallskip

\noindent
Benn, D. 2013, VSTAR data analysis software (http://www.aavso.org/node/803).

\smallskip

\noindent
Chiavassa, A., Haubois, X., Young, J.S., Plez, B., Josselin, E., Perrin, G.,
and Freytag, B., 2010, {\it Astron. Astrophys.}, {\bf 515}, 12.
 
\smallskip

\noindent
Fadeyev, Y.A. 2012, {\it Astron. Letters}, {\bf 38}, 260.

\smallskip

\noindent
Gray, D.F. 2000, {\it Astrophys. J.}, {\bf 532}, 487.

\smallskip

\noindent
Kervella, P., Le Bertre, T. and Perrin, G. (editors), 2013, {\it Betelgeuse
Workshop 2012: The Physics of Red Supergiants: Recent Advances and Open Questions}, EAS Publications Series, {\bf 60}.

\smallskip

\noindent
Kiss, L.L., Szab{\'o}, G.M. and Bedding, T.R. 2006, {\it Mon. Not. Roy. Astron. Soc.}, {\bf 372}, 1721.

\smallskip

\noindent
Nicholls, C.P., Wood, P.R., Cioni, M.-R.L. and Soszy{\'n}ski, I. 2009, {\it Mon. Not. Roy. Astron. Soc.}, {\bf 399}, 2063.

\smallskip

\noindent
Percy, J.R. and Sato, H., 2009, {\it J. Roy. Astron. Soc. Canada}, {\bf 103}, 11.

\smallskip

\noindent
Percy, J.R. and Abachi, R., 2013, {\it JAAVSO}, in press; on-line at:

\smallskip

http://www.aaavso.org/sites/default/files/jaavso/ej243.pdf

\smallskip

\noindent
Pugh, T. and Gray, D.F. 2013a, {\it Astron. J.}, {\bf 145}, 38.

\smallskip

\noindent
Pugh, T. and Gray, D.F. 2013b, {\it Astron. J.}, in press.

\smallskip

\noindent
Templeton, M.R., Mattei, J.A., and Willson, L.A., 2005, {\it Astron. J.},
{\bf 130}, 776.

\small

\begin{table}\small
\caption{Amplitude Variability of Pulsating Red Supergiants.} 
\begin{tabular}{rrrrrrr}
\hline
Star & P(d) & JD(1) & A Range & N & L(d) & L/P \\
\hline
SS And* & 159 & 2429500 & 0.15-0.55 & 9.5 & 2789 & 18 \\
VY CMa* & 1440 & 2440500 & 0.33-0.52 & 0.5 & 41000 & 26 \\
RT Car* & 448 & 2437000* & 0.07-0.31 & 2.5 & 6600 & 15 \\
RT Car* & 2060 & 2437000* & 0.07-0.11 & $<$0.5 & $>$39000 & $>$19 \\
BO Car* & 330 & 2439000 & 0.07-0.27 & 3.5 & 5000 & 15 \\
CL Car* & 500 & 2434800 & 0.08-0.43 & 0.5 & 43400 & 87 \\
EV Car* & 820 & 2440000 & 0.10-0.50 & 0.5 & 33000 & 40 \\
TZ Cas & 3100 & 2438000 & 0.11-0.16 & $<$0.2 & $>$86000 & $>$28 \\
PZ Cas & 840 & 2440000 & 0.13-0.50 & 0.5 & 33000 & 39 \\
$\mu$ Cep* & 870 & 2394500 & 0.07-0.26 & 5.0 & 11800 & 14 \\
$\mu$ Cep* & 4525 & 2394500 & 0.07-0.09 & 0.8 & 80000 & 18 \\
RW Cyg & 645 & 2416000 & 0.09-0.19 & 1.25 & 32400 & 50 \\
BC Cyg & 700 & 2439500 & 0.15-0.51 & 0.5 & 33000 & 46 \\
TV Gem* & 426 & 2431000 & 0.08-0.29 & 2.5 & 9600 & 23 \\
TV Gem* & 2570 & 2431000 & 0.16-0.21 & $<$0.2 & $>$96000 & $>$38 \\ 
$\alpha$ Her* & 1550 & 2425000 & 0.05-0.09 & 1.0 & 31500 & 20 \\
W Ind* & 198 & 2440000 & 0.23-0.42 & 3.75 & 2933 & 15 \\
$\alpha$ Ori* & 388 & 2420000 & 0.03-0.24 & 5.25 & 6000 & 15 \\
$\alpha$ Ori* & 2300 & 2420000 & 0.08-0.14 & 0.5 & 73000 & 32 \\
S Per* & 813 & 2419000 & 0.32-0.85 & 2.25 & 16670 & 21 \\
T Per & 2500 & 2410000 & 0.05-0.10 & $<$0.5 & $>$93000 & $>$37 \\
W Per* & 500 & 2415000 & 0.20-0.47 & 3.5 & 11857 & 24 \\
W Per* & 2875 & 2415000 & 0.17-0.19 & 1.0 & 41500 & 14 \\
SU Per & 450 & 2432500 & 0.07-0.22 & 2.5 & 21500 & 46 \\
SU Per & 3300 & 2432500 & 0.07-0.22 & 0.75 & 32000 & 10 \\
XX Per* & 2400 & 2422500 & 0.05-0.10 & $<$0.5 & $>$113000 & $>$47 \\
BU Per* & 381 & 2432500 & 0.13-0.15 & $<$0.25 & $>$66000 & $>$173 \\
VX Sgr & 754 & 2427500 & 0.55-1.35 & $<$1.0 & $>$29000 & $>$38 \\
AH Sco & 765 & 2440000 & 0.38-0.53 & 1.5 & 11667 & 16 \\
$\alpha$ Sco* & 1750 & 2431000 & 0.10-0.40 & 0.5 & 51000 & 29 \\
CE Tau & 1300 & 2435000 & 0.06-0.13 & 0.6 & 35833 & 28 \\
W Tri* & 680 & 2431500 & 0.08-0.11 & 1.0 & 25000 & 37 \\
\hline
\end{tabular}
\end{table}

\end{document}